\documentclass[twocolumn]{aastex631}

\usepackage{here}

\newcommand{\degdeg}{^{\circ}}

\newcommand{\sag}{Stars and Galaxies}

\submitjournal{ApJ}
\shortauthors{Beniyama et al.}
\graphicspath{{./}{figures/}}

\begin{document}

\title{Photometry and Polarimetry of 2010 XC$_{15}$:\\Observational Confirmation of E-type Near-Earth Asteroid Pair}
\shorttitle{Photometry and polarimetry of 2010 XC$_{15}$}

\correspondingauthor{Jin Beniyama}
\email{beniyama@ioa.s.u-tokyo.ac.jp}
\author[0000-0003-4863-5577]{Jin Beniyama}
\affiliation{%
Institute of Astronomy, Graduate School of Science,\\
The University of Tokyo, 2-21-1 Osawa, Mitaka, Tokyo 181-0015, Japan}
\affiliation{%
Department of Astronomy, Graduate School of Science,\\
The University of Tokyo, 7-3-1 Hongo, Bunkyo-ku, Tokyo 113-0033, Japan}
\author[0000-0002-8792-2205]{Shigeyuki Sako}
\affiliation{%
Institute of Astronomy, Graduate School of Science,\\
The University of Tokyo, 2-21-1 Osawa, Mitaka, Tokyo 181-0015, Japan}
\affiliation{%
UTokyo Organization for Planetary Space Science, 
The University of Tokyo, 7-3-1 Hongo, Bunkyo-ku, 
Tokyo 113-0033, Japan}
\affiliation{%
Collaborative Research Organization for
Space Science and Technology, 
The University of Tokyo, 7-3-1 Hongo, 
Bunkyo-ku, Tokyo 113-0033, Japan}

\author[0000-0001-6977-351X]{Katsuhito Ohtsuka}
\affiliation{%
Tokyo Meteor Network, 1–27–5 Daisawa, Setagaya-ku, Tokyo 155–0032, Japan}
\author[0000-0003-1726-6158]{Tomohiko Sekiguchi}
\affiliation{%
Asahikawa Campus, Hokkaido University of Education, 
Hokumon 9, Asahikawa, Hokkaido 070-8621, Japan}
\author[0000-0002-7332-2479]{Masateru Ishiguro}
\affiliation{%
Department of Physics and Astronomy, 
Seoul National University, 1 Gwanak-ro, Gwanak-gu, 
Seoul 08826, Republic of Korea}
\affiliation{%
SNU Astronomy Research Center, 
Seoul National University, 
1 Gwanak-ro, Gwanak-gu, Seoul 08826, Republic of Korea}
\author[0000-0002-7363-187X]{Daisuke Kuroda}
\author[0000-0001-7501-8983]{Seirato Urakawa}
\affiliation{%
Bisei Spaceguard Center, Japan Spaceguard Association, 
1716-3 Okura, Bisei, Ibara, Okayama 714-1411, Japan}
\author[0000-0002-3286-911X]{Fumi Yoshida}
\affiliation{%
School of Medicine, Department of Basic Sciences,
University of Occupational and Environmental Health, 1-1 Iseigaoka,
Yahata, Kitakyusyu, Fukuoka 807-8555, Japan}
\affiliation{%
Planetary Exploration Research Center,
Chiba Institute of Technology, 2–17–1 Tsudanuma, Narashino,
Chiba 275–0016, Japan}

\author{Asami Takumi}
\affiliation{%
Open University of Japan, 2-11 Wakaba, Mihama-ku, Chiba 261-8586, Japan}
\affiliation{%
National Astronomical Observatory of Japan, 
2-21-1 Osawa, Mitaka, Tokyo 181-8588, Japan}
\author[0000-0002-6874-5178]{Natsuho Maeda}
\affiliation{%
Department of Planetology, Kobe University, 1-1 Rokkodai-cho, Nada-ku, 
Kobe 657-8501, Japan}
\author{Jun Takahashi}
\affiliation{%
Center for Astronomy, University of Hyogo, 407-2 Nishigaichi, Sayo, Hyogo 679-5313, Japan}
\author[0000-0002-7084-0860]{Seiko Takagi}
\author{Hiroaki Saito}
\affiliation{%
Department of Earth and Planetary Sciences, Faculty of Science, Hokkaido University,
Kita-ku, Sapporo, Hokkaido 060-0810, Japan}
\author{Tatsuya Nakaoka}
\affiliation{%
Hiroshima Astrophysical Science Center, Hiroshima University, 1-3-1 Kagamiyama, Higashi-Hiroshima,
Hiroshima 739-8526, Japan}
\author{Tomoki Saito}
\author{\textbf{Tomohiro Ohshima}}
\affiliation{%
Center for Astronomy, University of Hyogo, 407-2 Nishigaichi, Sayo, Hyogo 679-5313, Japan}
\author[0000-0002-0643-7946]{Ryo Imazawa}
\affiliation{%
Department of Physics, Graduate School of Advanced Science and Engineering, Hiroshima University, 1-3-1 Kagamiyama, 
Higashi-Hiroshima, Hiroshima 739-8526, Japan}

\author[0000-0002-4115-7122]{Masato Kagitani}
\affiliation{%
Planetary Plasma and Atmospheric Research Center, Graduate School of Science, Tohoku University, 6-3 Aramaki-Aoba, Aoba-ku, Sendai, Miyagi 980-8578, Japan}
\author{Satoshi Takita}
\affiliation{%
Institute of Astronomy, Graduate School of Science,\\
The University of Tokyo, 2-21-1 Osawa, Mitaka, Tokyo 181-0015, Japan}

\begin{abstract}
Asteroid systems such as binaries and pairs are indicative of physical properties and 
dynamical histories of the Small Solar System Bodies. 
Although numerous observational and theoretical studies have been carried out, 
the formation mechanism of asteroid pairs is still unclear, especially for 
near-Earth asteroid (NEA) pairs.
We conducted a series of optical photometric and polarimetric observations of a small NEA 2010 XC$_{15}$ in 2022 December to investigate its surface properties.
The rotation period of 2010 XC$_{15}$ is possibly a few to several dozen hours and
color indices of 2010 XC$_{15}$ are derived as $g-r=0.435\pm0.008$, $r-i=0.158\pm0.017$, and $r-z=0.186\pm0.009$ in the Pan-STARRS system.
The linear polarization degrees of 2010 XC$_{15}$ are a few percent 
at the phase angle range of 58$^{\circ}$ to 114$^{\circ}$.
We found that 2010 XC$_{15}$ is a rare E-type NEA on the basis of its photometric and polarimetric properties. 
Taking the similarity of not only physical properties but also dynamical integrals and the rarity of E-type NEAs into account,
we suppose that 2010 XC$_{15}$ and 1998 WT$_{24}$ are of common origin (i.e., asteroid pair).
These two NEAs are the sixth NEA pair and first E-type NEA pair ever confirmed, possibly formed by rotational fission.
We conjecture that the parent body of 2010 XC$_{15}$ and 1998 WT$_{24}$ was 
transported from the main-belt through the $\nu_6$ resonance or Hungaria region.
\end{abstract}

\section{Introduction}\label{sec:intro}
Gravitationally bound multiple systems in the Small Solar System Bodies (SSSBs) 
have an important role to study the Solar System;
they provide us with rare opportunities to understand the physical properties, such as density and mass of the SSSBs
\citep[e.g.,][for reviews]{Merline2002, Margot2015, Walsh2015}.
The usefulness of the binary system has been demonstrated by a planetary defense mission, 
the Double Asteroid Redirection Test \citep[DART,][]{Rivkin2021}.
The spacecraft changed the orbital period of Dimorphos (secondary) around Didymos (primary) by
approximately 33 min \citep{Thomas2023}, which indicates a large momentum transfer efficiency.
Additional characterizations will be performed by Hera, a European Space Agency rendezvous mission
\citep{Michel2022}.
The DART mission provided never before heard of science cases by comparing the observational results 
pre- and post- DART impact \citep[e.g.,][]{Bagnulo2023}.

The first satellite of SSSBs was discovered around the main-belt asteroid (MBA) (243) Ida 
by the Galileo spacecraft \citep{Chapman1995, Belton1995}.
Later, lots of binaries were identified by 
lightcurve observations \citep[e.g.,][]{Pravec2007}
and 
direct imaging \citep[e.g.,][]{Rojo2011, Broz2022}.
A number of binary systems have been reported in the near-Earth region as well.
Binary asteroids are not a rare population in near-Earth asteroids \citep[NEAs,][]{Pravec2006, Scheirich2009}.
The lightcurve observations are, in principle, biased towards the detection of close binaries
while radar observations are biased towards the detection of binaries with large separations \citep{Margot2002, Ostro2006}.
Homogeneous spectra of binaries have been reported by the 
simultaneous spectroscopic and photometric observations \citep{Polishook2009}.
Besides binary systems, 
\cite{Marchis2005} reported
the MBA (87) Sylvia as the first triple system observed using the
Adaptive Optics (AO) system on the Very Large Telescope (VLT).
Recently, \cite{Berdeu2022} discovered the third satellite around the MBA (130) Elektra
using the Spectro-Polarimetric High-contrast Exoplanet REsearch adaptive optics 
system and coronagraphic facility (SPHERE) and an integral field spectrograph (IFS)
on the VLT, making Elektra the first quadruple asteroid.

Gravitationally unbound 
two asteroid components
with common origin in the SSSBs are called pairs.
Asteroid pairs were first reported by \cite{Vokrouhlicky2008}.
The formation time of the pairs (i.e., ages) were estimated with
differences in mean anomaly $\Delta M$ and semimajor axis separation $\Delta a$ of cluster members
\citep{Vokrouhlicky2008}.
As with the binaries, lots of lightcurve \citep{Pravec2019}
and spectroscopic observations \citep{Duddy2013, Polishook2014a, Polishook2014c} have been carried out.
Most asteroid pairs have similar spectra \citep{Polishook2014a, Polishook2014c}
while the pair (17198) Gorjup and (229056) 2004 FC$_{126}$ 
have slightly different spectra \citep{Duddy2013}.
Asteroid pairs are mainly studied for MBAs since in the near-Earth region, 
only four asteroid pairs have currently been reported.

The first NEA pair\footnote{We use the term NEA pair regardless of whether the age of the pair is estimated or not.},
(3200) Phaethon and (155140) 2005 UD,
was discovered by \cite{Ohtsuka2006}.
They performed backward and forward orbital integrations 
of both NEAs and found that their same orbital evolutionary phase has been shifted by 4600 years,
which indicates they are of common origin.
The similarity of the visible spectra of Phaethon and 2005 UD classified as B-types supports the hypothesis of common origin.
Their similar polarimetric properties also support this hypothesis \citep{Devogele2020, Ishiguro2022};
however, their near-infrared spectra are different \citep{Kareta2021}.

\cite{Ohtsuka2008a} reported that another NEA (225416) 1999 YC is dynamically related to Phaethon and 2005 UD.
The visible color measurements of 1999 YC indicate that the NEA is 
classified as C-type taxonomy, not B-type one \citep{Kasuga2008}.  
\cite{Ohtsuka2007} found the second NEA pair,
(1566) Icarus--2007 MK$_6$, using the same method as \cite{Ohtsuka2006}.
As written in \cite{Ohtsuka2007},
the determination of additional physical parameters of both Icarus and 2007 MK$_6$ is crucial to further investigate their common origin.
After more than ten years, \cite{Marcos2019c} reported the discovery of 
a third NEA pair 2017 SN$_{16}$--2018 RY$_7$. 
The orbits of the pair asteroids are stable owing to the 3:5 mean motion resonance (MMR) with Venus, avoiding close encounters with it.
\cite{Moskovitz2019} found that 
2017 SN$_{16}$ and 2018 RY$_7$ have similar visible spectra,
which supports that they are of common origin.
\cite{Moskovitz2019} also found that 
the visible spectra of the two NEAs 2015 EE$_7$ and 2015 FP$_{124}$ resemble each other
and concluded that they are a pair candidate.
Recently, \cite{Fatka2022} discovered a very young NEA pair 2019 PR$_2$--2019 QR$_6$.
The visible spectra of 2019 RP$_2$ and 2019 QR$_6$ are similar to primitive D-types, 
which is a rare type in the near-Earth region.
The D-types have similar colors with cometary nuclei \citep[e.g.,][]{Capaccioni2015}.
Their backward orbital integrations did not show the close encounters of the two NEAs (i.e., break up event) without cometary-like non-gravitational force.
They concluded that the separation time is approximately 300 years ago with the cometary-like non-gravitational model, 
which implies that 2019 PR$_2$--2019 QR$_6$ is the youngest pair known to date. 

Some formation mechanisms have been proposed for the multiple asteroid systems,
of which one of the leading formation mechanisms 
is the rotational fission \citep{Walsh2008, Jacobson2011, Jacobson2014, Jacobson2016}.
We use the term rotational fission in this paper instead of rotational breakup or rotational disruption.
\cite{Walsh2008} found that satellites are formed by mass shedding events
after a spin-up by the Yarkovsky-O'Keefe-Radzievskii-Paddack (YORP) effect,
which arises from the asymmetry of scattered sunlight and thermal radiation from its surface \citep{Rubincam2000}.
The rotational fission is consistent with the observational fact that 
many primaries of binary and pair systems are fast-rotating asteroids \citep{Pravec2010, Pravec2019}.
As for the NEA pairs, the fast rotation of the primary Icarus approximately 2.3 hr leverages the rotational fission hypothesis.
In genenal, however, the dynamical environments of MBAs and NEAs are different;
MBAs are relatively stable in terms of dynamics, whereas NEAs are in chaotic nature due to frequent close encounters with the planets.
Confirmed NEA pairs are relatively free from close approaches with the inner planets 
owing to such as large eccentricity, large inclination, and MMR.
Thus, the formation mechanisms of NEA pairs are still unclear, 
and other formation mechanisms such as tidal interaction with the planets are under considerations \citep{Richardson1998, Walsh2006, Scheeres2000}.
Additional observations of multiple systems 
are essential to reach a consensus regarding the origins of binary and pair systems.

The target NEA of this study, 2010 XC$_{15}$, was discovered by the Catalina Sky Survey \citep{Drake2009} with a 0.7~m Schmidt telescope on Mt. Bigelow on 2010 December 5.
Assuming an absolute magnitude $H$ of 21.70 and a geometric albedo
$p_\mathrm{V}$ of $0.350^{+0.176}_{-0.151}$, 
derived from thermal observations using 
the Infrared Array Camera (IRAC) on the Spizer Space Telescope (SST) in
2017 as part of the NEO Legacy project \footnote{\url{http://nearearthobjects.nau.edu/spitzerneos.html}}, 
the diameter of 2010 XC$_{15}$ was derived to be approximately 100~m.
The orbital elements of 2010 XC$_{15}$ 
resemble those of (33342) 1998 WT$_{24}$, which is a well-characterized E-type NEA \citep{Kiselev2002, Harris2007, Busch2008}.
1998 WT$_{24}$ has an effective diameter of $415\pm40$~m \citep{Busch2008}.
E-types are thought to have mineralogical links to the enstatite achondrite meteorites (Aubrites) 
composed of almost iron-free enstatite \citep{Zellner1975, Zellner1977}.
The orbital similarity criterion, $D_{\rm SH}$ \citep{Southworth1963} between two orbits, 
is as small as 0.04 presently, which is comparable to the well-established Phaethon--Geminid meteor stream relation.
This value is smaller than the empirical cutoff for significance, $\sim$0.20 \citep{Drummond1991}, 
and indicates the orbital similarity of 2010 XC$_{15}$ and 1998 WT$_{24}$.
The apparent magnitude of 2010 XC$_{15}$ in the $V$-band was brightened up to 13 mag in 2022 December,
which allowed us to constrain the surface properties of a small asteroid.
We conducted a one-week observation campaign of 2010 XC$_{15}$ with multiple telescopes in 2022 December.
In section 2, we describe our methods: photometry, polarimetry, and orbital integrations.
The results are presented in section 3.
The surface properties of 2010 XC$_{15}$ are investigated in section 4.
The possible dynamical history and origin of 2010 XC$_{15}$ and 1998 WT$_{24}$ are also discussed.

\section{Methods}
\subsection{Photometric observations} \label{subsec:phot}
We performed multicolor photometry of 2010 XC$_{15}$
using the TriColor CMOS Camera and Spectrograph (TriCCS) on the Seimei 3.8 m telescope \citep{Kurita2020}
at the Kyoto University Okayama Observatory (133.5967$\degdeg$E, 34.5769$\degdeg$N, and 355 m in altitude).
The details of the photometry are presented in Table \ref{tab:phot}.
The single exposure time was set to 5.0 s,
and the telescope was operated in the non-sidereal tracking mode.
The data simultaneously obtained with $g$, $r$, and $i$ or $z$-bands in the Pan-STARRS system \citep{Chambers2016} 
were analyzed using the same procedure as described in \cite{Beniyama2023a}.

After bias subtraction, dark subtraction, and flat-fielding,
the astrometry of all the reduced images was performed using \texttt{astrometry.net} software \citep{Lang2010}.
Cosmic ray related signals were removed using the \texttt{Python} package \texttt{Astro-SCRAPPY} \citep{McCully2018}
based on Pieter van Dokkum's \texttt{L.A.Cosmic} algorithm \citep{vanDokkum2001}.
The circular aperture photometry was performed on 2010 XC$_{15}$ 
and reference stars in each frame with the SExtractor-based \texttt{Python} package \texttt{SEP}
\citep{Bertin1996, Barbary2015}.
The aperture radii were set to twice the size of the FWHMs of the PSFs of reference stars.
The light-travel time of the target asteroid was corrected 
to obtain the time-series colors and magnitudes \citep{Harris1989}.

Reference stars meeting any of the criterion below were not used in the determination of colors and magnitudes:
uncertainties in $g$, $r$, $i$, or $z$-band magnitudes 
in Pan-STARRS Data Release 2 \citep[DR2,][]{Flewelling2020} are larger than 0.05,
$(g-r)_\mathrm{PS}> 1.1$, 
$(g-r)_\mathrm{PS}< 0.0$, 
$(r-i)_\mathrm{PS}> 0.8$,
or 
$(r-i)_\mathrm{PS}< 0.0$, where
$(g - r)_{\mathrm{PS}}$ and $(r - i)_{\mathrm{PS}}$ are colors in Pan-STARRS system.
We discarded the sources close to the edges of the image frame 
(100 pixels from the edge) or with any other sources within the aperture.
Objects categorized as extended sources, possible quasars, and variable stars were removed 
with the \texttt{objinfoflag} and \texttt{objfilterflag} in Pan-STARRS DR2.
After deriving the colors and magnitudes of 2010 XC$_{15}$ in the Pan-STARRS system for each frame,
we used a binning of 60~s for all magnitudes and colors.

\begin{deluxetable*}{cccrcccccc}
        \tablenum{1}
        \tablecaption{Summary of photometric observations\label{tab:phot}}
        \tablewidth{0pt}
        \tablehead{
            \colhead{Obs. Date} & Filters & \colhead{$T_\mathrm{exp}$} & \colhead{$N_\mathrm{exp}$} & $V$ & $\alpha$ & $\Delta$ & $r_\mathrm{h}$ & Air Mass & Weather \\
            \colhead{(UT)}     &               & \colhead{(s)}              &                            & (mag) & (deg) & (au) & (au) &  &
        }
        \decimals
        \startdata
        2022 Dec 22 16:25:41--18:11:52& $g,r,i$ &5 & 478 & 15.9 & 58.5 &0.030 & 0.999 & 1.24--1.52 & Cirrus\\
2022 Dec 23 16:14:26--21:11:10& $g,r,i$ &5 & 1699 & 15.5 & 59.2 &0.024 & 0.996 & 1.20--1.57 & Clear\\
2022 Dec 23 18:35:51--19:00:50& $g,r,z$ &5 & 172 & 15.5 & 59.2 &0.024 & 0.996 & 1.20--1.21 & Clear\\
2022 Dec 24 15:45:24--21:08:32& $g,r,i$ &5 & 1230 & 14.9 & 60.9 &0.018 & 0.992 & 1.17--1.80 & Clear\\
2022 Dec 25 18:18:53--21:04:21& $g,r,z$ &5 & 821 & 14.2 & 65.0 &0.012 & 0.989 & 1.13--1.22 & Clear\\
  \enddata
            \tablecomments{
            Observation time in UT in mid-time of exposure (Obs. Date), filters (Filters), 
            exposure time ($T_{\mathrm{exp}}$),
            the number of exposures ($N_\mathrm{exp}$),
            and weather condition (Weather) are listed.
            Predicted $V$-band apparent magnitude ($V$), 
            phase angle ($\alpha$),
            distance between 2010 XC$_{15}$ and observer ($\Delta$),
            and 
            distance between 
            2010 XC$_{15}$ and the Sun ($r_\mathrm{h}$) at the observation starting time
            are from NASA Jet Propulsion Laboratory (JPL) HORIZONS
             as of 2023 May 11 (UTC).
            Elevations to calculate air mass range (Air Mass) are 
            also from NASA JPL HORIZONS.
            }
            \end{deluxetable*}

\subsection{Polarimetric observations} \label{subsec:pol}
We conducted polarimetric observations at three sites in Japan;
Nayoro Observatory 
(142.4830$\degdeg$E, 44.3736$\degdeg$N, and 151 m in altitude, the Minor Planet Center code Q33, 
hereinafter referred to as NO),
Nishi–Harima Astronomical Observatory 
(134.3356$\degdeg$E, 35.0253$\degdeg$N, and 449 m in altitude, 
hereinafter referred to as NHAO),
and Higashi-Hiroshima Observatory (132.7767$\degdeg$E, 34.3775$\degdeg$N, and 511.2 m in altitude, 
hereinafter referred to as HHO).
The observing specifications of polarimetry are summarized in Table \ref{tab:pol}.
We used the Multi-Spectral Imager \citep[MSI,][]{Watanabe2012} mounted on 1.6 m Pirka Telescope at NO,
the Wide Field Grism Spectrograph 2
\citep[WFGS2,][]{Uehara2004, Kawakami2021} mounted on 2.0 m Nayuta Telescope at NHAO,
and the Hiroshima Optical and Near-InfraRed Camera
\citep[HONIR, ][]{Akitaya2014} mounted on 1.5 m Kanata telescope at HHO.
Wollaston prisms and rotatable half-wave plates are installed in all the three instruments;
thus, data obtained at the three sites were 
analyzed in the same standard reduction procedure \citep[e.g.,][]{Kawabata1999, Ishiguro2017}.
We used the SExtractor-based \texttt{Python} package \texttt{SEP}
for the circular aperture photometry.
We derived linear polarization degrees relative 
to the perpendicular to the scattering plane, $P_\mathrm{r}$, 
and position angles of polarization, $\theta_\mathrm{r}$.
Additionally, we observed a polarimetric standard star HD 19820 \citep{Schmidt1992} 
to verify the consistency of our measurements (Appendix \ref{app:polstan}).
We considered the deviations between the polarimetric parameters 
in the literature and those derived here 
as systematic uncertainties in the measurements of the polarimetric parameters of 2010 XC$_{15}$.

\begin{deluxetable*}{ccccccccccc}
        \tablenum{2}
        \tablecaption{Summary of polarimetric observations\label{tab:pol}}
        \tablewidth{0pt}
        \tablehead{
            \colhead{Obs. Date} & Inst. & Filter & \colhead{$T_\mathrm{exp}$} & \colhead{$N_\mathrm{exp}$} & $V$ & $\alpha$   & $\phi$      & $P_\mathrm{r}$     & $\theta_\mathrm{r}$    & Air Mass \\
            \colhead{(UT)}     &       &        & \colhead{(s)}              &                   & (mag) & ($^\circ$) &  ($^\circ$) & (\%)   & ($^\circ$) &    
        }
        \decimals
        \startdata
        2022 Dec 20 15:30:37--16:29:49& WFGS2 & $R_\mathrm{C}$ & 300 & 8 & 16.7 & 58.2 & 298.2 &$1.61\pm0.44$ & $-7.43\pm7.85$ & 1.51--1.91 \\
2022 Dec 21 15:46:40--16:37:10& MSI & $R_\mathrm{C}$ & 180 & 12 & 16.3 & 58.2 & 297.9 &$1.69\pm0.56$ & $-2.00\pm8.51$ & 1.57--1.79 \\
2022 Dec 24 20:36:51--20:55:13& HONIR & $R_\mathrm{C}$ & 115 & 8 & 14.9 & 61.0 & 294.9 &$1.36\pm0.20$ & $1.09\pm4.09$ & 1.25--1.29 \\
2022 Dec 25 16:57:21--21:12:51& HONIR & $R_\mathrm{C}$ & 115 & 16 & 14.2 & 64.9 & 292.7 &$1.27\pm0.17$ & $4.26\pm3.56$ & 1.13--1.42 \\
2022 Dec 26 17:40:04--18:49:09& HONIR & $R_\mathrm{C}$ & 115 & 16 & 13.5 & 75.8 & 291.0 &$1.74\pm0.08$ & $1.44\pm1.85$ & 1.17--1.38 \\
2022 Dec 27 19:36:23--20:57:28& WFGS2 & $R_\mathrm{C}$ & 60 & 32 & 14.4 & 113.1 & 316.7 &$1.85\pm0.25$ & $-7.41\pm2.57$ & 1.36--1.78 \\
2022 Dec 27 20:20:18--21:30:02& HONIR & $R_\mathrm{C}$ & 115 & 8 & 14.4 & 114.2 & 318.1 &$1.82\pm0.15$ & $-0.43\pm2.45$ & 1.28--1.56 \\
  \enddata
            \tablecomments{
            Observation time in UT in mid-time of exposure (Obs. Date), instument (Inst.), filter (Filter), 
            exposure time ($T_{\mathrm{exp}}$),
            and the number of exposures ($N_\mathrm{exp}$) are listed.
            Predicted $V$-band apparent magnitude ($V$),
            phase angle ($\alpha$), and 
            the position angle of the scattering plane ($\phi$)
            are from NASA JPL HORIZONS
             as of 2023 May 11 (UTC).
            Elevations to calculate air mass range (Air Mass) are 
            also from NASA JPL HORIZONS.
            }
            \end{deluxetable*}

\subsection{Orbital integrations}
Orbits of 2010 XC$_{15}$ and 
a well-characterized E-type NEA 1998 WT$_{24}$
resemble each other with a small $D_\mathrm{SH}$ of 0.04.
The geometric albedos of these two NEAs derived in previous studies 
are in agreement \citep{Kiselev2002}.
This suggests that the two NEAs could be of common origin and 
it makes sense to investigate the orbital history of these two NEAs.

To investigate the dynamical link and origins of 2010 XC$_{15}$ and 1998 WT$_{24}$,
we performed backward orbital integrations using \texttt{Mercury 6},
a general purpose software package for problems in solar system dynamics \citep{Chambers1997}.
We computed close encounters accurately with the general Bulirsch-Stoer algorithm available in \texttt{Mercury 6}.
We also performed orbital integrations with the hybrid of symplectic and Bulirsch-Stoer integrator in \texttt{Mercury 6}
and confirmed the results of our integrations do not significantly change.

The nominal orbital elements of asteroids, their uncertainties, and the covariance data were referred to NASA JPL Small-Body DataBase (SBDB)
\footnote{https://ssd-api.jpl.nasa.gov/doc/sbdb.html}.
The coordinates and velocities of the Sun and eight planets were obtained from NASA JPL HORIZONS.
The non-gravitational transverse acceleration parameters, $A_2$ \citep{Farnocchia2013}, are given for
both 2010 XC$_{15}$ and 1998 WT$_{24}$ in NASA JPL SBDB.
The semimajor axes of these two asteroids have shrunk, possibly due to the Yarkovsky effect
\citep[e.g.,][]{Vokrouhlicky1998, Vokrouhlicky2000, Bottke2006}.
Thus, we also considered the non-gravitational transverse acceleration by 
setting $A_2$ parameters in the integrations.
We investigated time evolutions of the asteroids under 
the gravity of the Sun and eight planets in the solar system.
The first time step was set to 0.1 days in all integrations.
The coordinates and velocities of asteroids were output every 300 days.
We converted coordinates and velocities to orbital elements using \texttt{element6}, a program in \texttt{Mercury 6}.

Orbital evolution of NEAs often turns chaotic after a short period ($\sim$ 100 years) 
of integration due to frequent close encounters with planets \citep{Yoshikawa2000}.
We generated clones utilizing the classical Monte Carlo using the Covariance Matrix 
\citep[MCCM, ][]{Avdyushev2007, Marcos2015}
approach to check this chaotic behavior.
Each clone had initial orbital elements slightly different from the nominal ones.
In total, we generated 1000 clones considering uncertainties of 
eccentricity ($e$) , 
perihelion distance ($q$),
time of perihelion passage ($\tau$),
longitude of ascending node ($\Omega$),
argument of perihelion ($\omega$), 
and inclination ($i$).
The 1000 clones were generated with random numbers made with the 
\texttt{np.random.randn} function in the \texttt{Python} package \texttt{NumPy} 
\citep{Oliphant2015, Harris2020}. 
We set the seed of the random number as 0.
The semimajor axis ($a$) was calculated as $a = q/(1-e)$ after the orbital integrations.
The epochs of orbital elements of 2010 XC$_{15}$ and 1998 WT$_{24}$ were
2018 January 1 UT, JD 2458119.5, and 2016 April 16 UT, JD 2457494.5,
respectively.

To discuss whether or not the asteroids are of common origin, 
we used the following three integrals of motion of the asteroids 
in the circular restricted three-body problem \citep{Lidov1962}
in the von Zeipel-Lidov-Kozai (vZLK) oscillation \citep{Lidov1962, Kozai1962, Ito2019}:
\begin{eqnarray}
    C_0 &=& \frac{1}{a} \equiv \mathrm{const} ,\\
    C_1 &=&  (1-e^2)\cos^2i \equiv \mathrm{const} ,\\
    C_2 &=&  e^2(0.4-\sin^2 i \sin^2\omega) \equiv \mathrm{const}.
\end{eqnarray}
When an asteroid breaks up into multiple bodies by catastrophic impacts or rotational fission,
the fragments have slightly different orbital elements at that time.
In the orbital evolutions of NEAs, their orbital parameters could be drastically changed due to close encounters with planets.
On the other hand, the integrals of $C_0$, $C_1$, and $C_2$ are kept constant for a longer time.
Thus, these integrals should be good indicators to determine whether the asteroids are an NEA pair \citep{Ohtsuka2006, Ohtsuka2007}.

\section{Results} \label{sec:res}
\subsection{Lightcurves and rotation period}
\begin{figure*}[ht]
\plotone{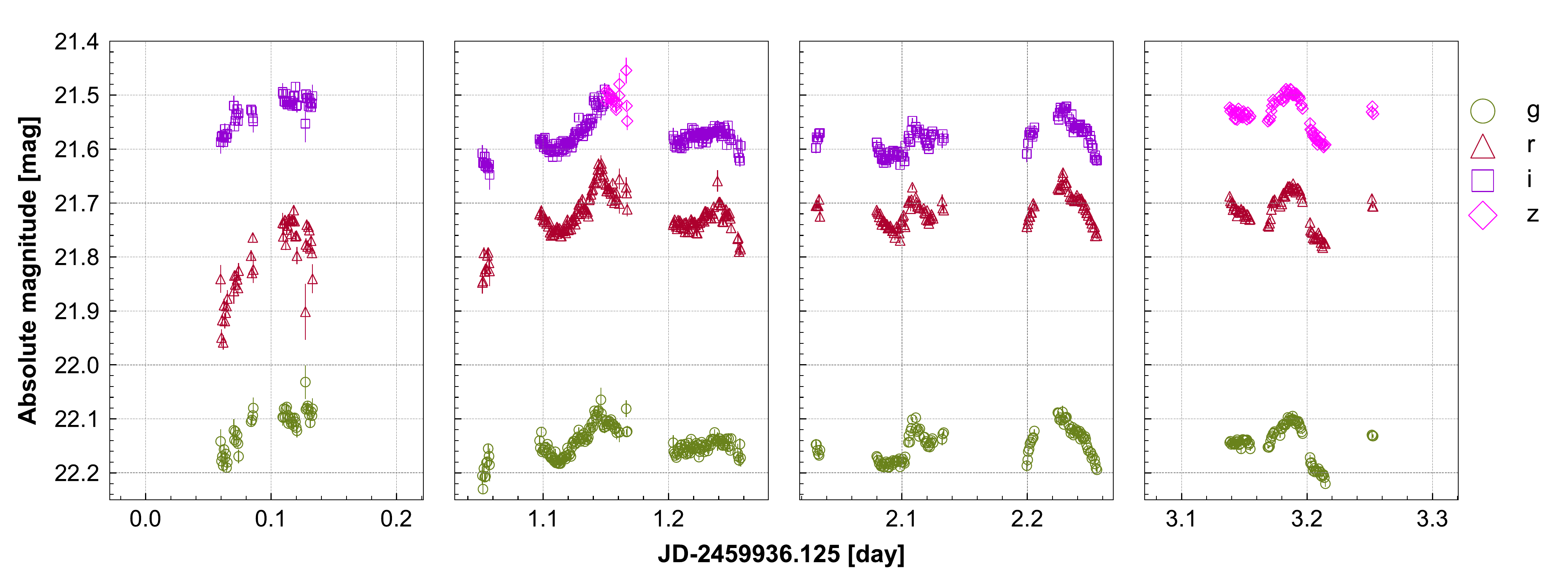}
\caption{
Phase corrected lightcurves of 2010 XC$_{15}$.
Phase corrected lightcurves in the 
$g$- (circles), $r$- (triangles), $i$- (squares), and $z$-bands (diamonds)
are presented.
Time zero is set to JD 2459936.125 (2022 December 22 15:00:00 UT).
Bars indicate the 1$\sigma$ uncertainties.
}
\label{fig:lc}
\end{figure*}

For periodic analysis, we calculated reduced magnitudes from observed magnitudes as:
\begin{equation}
    m_{\mathrm{red}, n}(\alpha) = m_{\mathrm{obs}, n}(\alpha) - 5\log_{10}{(\Delta r_\mathrm{h})},
\end{equation}
where $n$ is the index of the band, 
$m_{\mathrm{red}, n}(\alpha)$ 
is a reduced magnitude in $n$-band, 
$m_{\mathrm{obs}, n}(\alpha)$ 
is an observed magnitude in $n$-band, 
$\Delta$ is a distance between 2010 XC$_{15}$ and the Earth,
and  $r_\mathrm{h}$ is a distance between 2010 XC$_{15}$ and the Sun.
We performed the phase angle correction 
since the phase angle of 2010 XC$_{15}$ changed significantly during our observations, $\sim 6.5\degdeg$ 
(Table \ref{tab:phot}).
In general, 
an empirical relation,
such as the $H$-$G$ model \citep{Bowell1989}, 
which is originally applied for photometric data taken at phase angles smaller than $\sim30\degdeg$,
is used for the phase correction.
The phase angles of 2010 XC$_{15}$ in our observations are as large as 60$\degdeg$,
where the brightness dependence on the solar phase angle, phase curve, is still poorly understood. 
We assumed the phase curve in linear form as follows:
\begin{equation}
    m_{\mathrm{red}, n}(\alpha) = H_n + b\alpha, \label{eq:linear}
\end{equation}
where $H_n$ is an absolute magnitude in $n$-band,
and $b$ is a linear slope of the phase curve.
We converted the reduced magnitudes to absolute magnitudes using the Equation (\ref{eq:linear}). 
Carefully checking the corrected lightcurves while setting 
$b=0.010, 0.015, 0.020, 0.025, 0.030$, and 0.035 mag/deg,
we finally adopted $b=0.030$~mag deg$^{-1}$ for the corrected lightcurves, as shown in Figure \ref{fig:lc}.
The clear variations with amplitudes of approximately 0.1 mag are seen in the corrected lightcurves.

A rotation period of 2010 XC$_{15}$ has been reported in the Asteroid Lightcurve Database \citep[LCDB, ][]{Warner2009}.
The period derived as $2.673\pm0.001$~hr with two days of optical photometry in 2022 December by Petr Pravec
\footnote{\url{https://www.asu.cas.cz/~ppravec/newres.txt}} is a possible solution
with a quality code $U$ in the LCDB of 2-.
Another period has been derived from the radar observations using the Goldstone Radar at DSS-14
\footnote{\url{https://echo.jpl.nasa.gov/asteroids/2010XC15/2010XC15.2022.goldstone.planning.html}}.
Assuming that the effective diameter of 2010 XC$_{15}$ is approximately 150~m,
they concluded that the Doppler broadening approximately 1.3~Hz at a wavelength of 3.5~cm corresponds to a rotation period of approximately 11.5~hr.
The rotation period is \textbf{constrained}
using the relation given by 
$B=(4\pi D)/(\lambda P_\mathrm{rot}) \sin\beta$,
where $B$ is a measured bandwidth, $D$ is a diameter, $\lambda$ is a transmitted wavelength,
$P_\mathrm{rot}$ is a rotation period, and $\beta$ is the angle
between the line of sight and a spin vector of the object \citep{Ostro1988}.
The 11.5~hr can be derived assuming $\sin\beta$ of 1.
Thus, the 11.5~hr is an upper limit of the rotation period since the pole orientation of 
2010 XC$_{15}$ is unknown.

\begin{figure}[ht]
\plotone{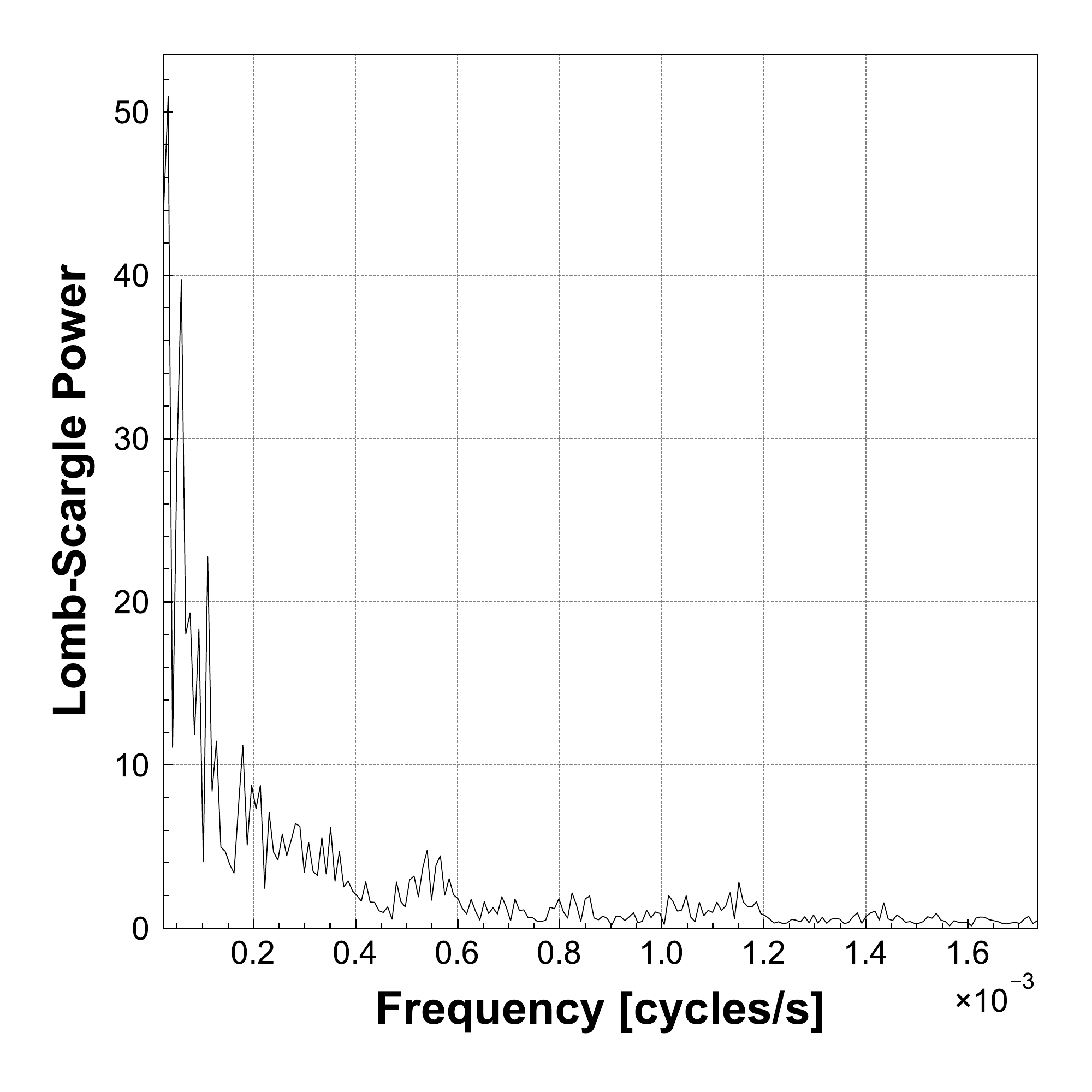}
\caption{
Lomb-Scargle periodogram of 2010 XC$_{15}$.
The number of harmonics of the model curve is five.
}
\label{fig:periodogram}
\end{figure}

\begin{figure}[ht]
\plotone{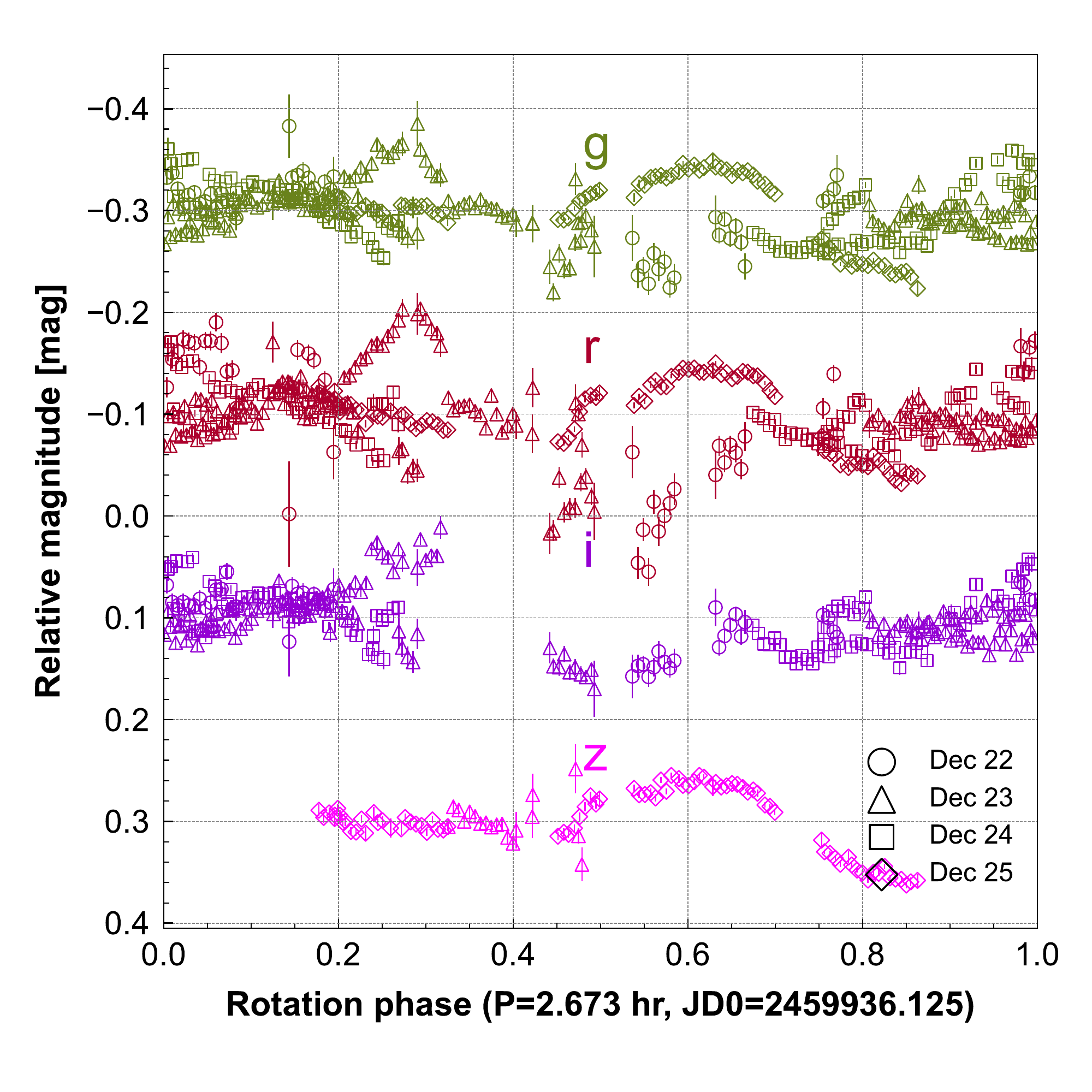}
\caption{
Phased lightcurves of 2010 XC$_{15}$.
From top to bottom, $g$-, $r$-, $i$-, and $z$-bands lightcurves are presented.
Lightcurves are folded by the derived period of 2.673~hr.
Phase zero is set to JD 2459936.125 (2022 December 22 15:00:00 UT).
Lightcurves in each band are horizontally offset by 0.2 mag for the sake of clarity.
Bars indicate the 1$\sigma$ uncertainties.
}
\label{fig:plc}
\end{figure}

We performed a periodic analysis with $g$-band lightcurves
using the Lomb-Scargle method limiting the period range to between 
0.16 and 11.5~hr, as shown in Figure \ref{fig:periodogram}.
We found no significant peak in the Lomb-Scargle periodogram.
Figure \ref{fig:plc} shows the phased lightcurves folded by a rotation period of 2.673~hr
reported in the LCDB.
We checked other phased lightcurves but it is difficult to conclude which peak
to prefer over the others as our lightcurve coverage is insufficient and a lightcurve amplitude of 2010 XC$_{15}$ is small
at approximately 0.1 mag.
The small lightcurve amplitude implies that 2010 XC$_{15}$ has a nearly spherical shape
\textbf{or its rotational axis is parallel to the line of sight}.
Lightcurves with small amplitudes sometimes 
show more than two maxima in a single rotation \citep{Harris2014}.
For example, (5404) Uemura shows six maxima \citep{Harris2014}, 
and (101955) Bennu shows three maxima \citep{Hergenrother2013} in a single rotation.
Additional observations are necessary for more constraint on the rotation state of 2010 XC$_{15}$.
Furthermore, we cannot rule out the possibility that 2010 XC$_{15}$ is a non-principal axis rotator 
\citep[i.e., tumbler,][]{Paolicchi2002, Pravec2005}.

\subsection{Colors and reflectance spectrum}
\begin{figure*}[ht]
\plotone{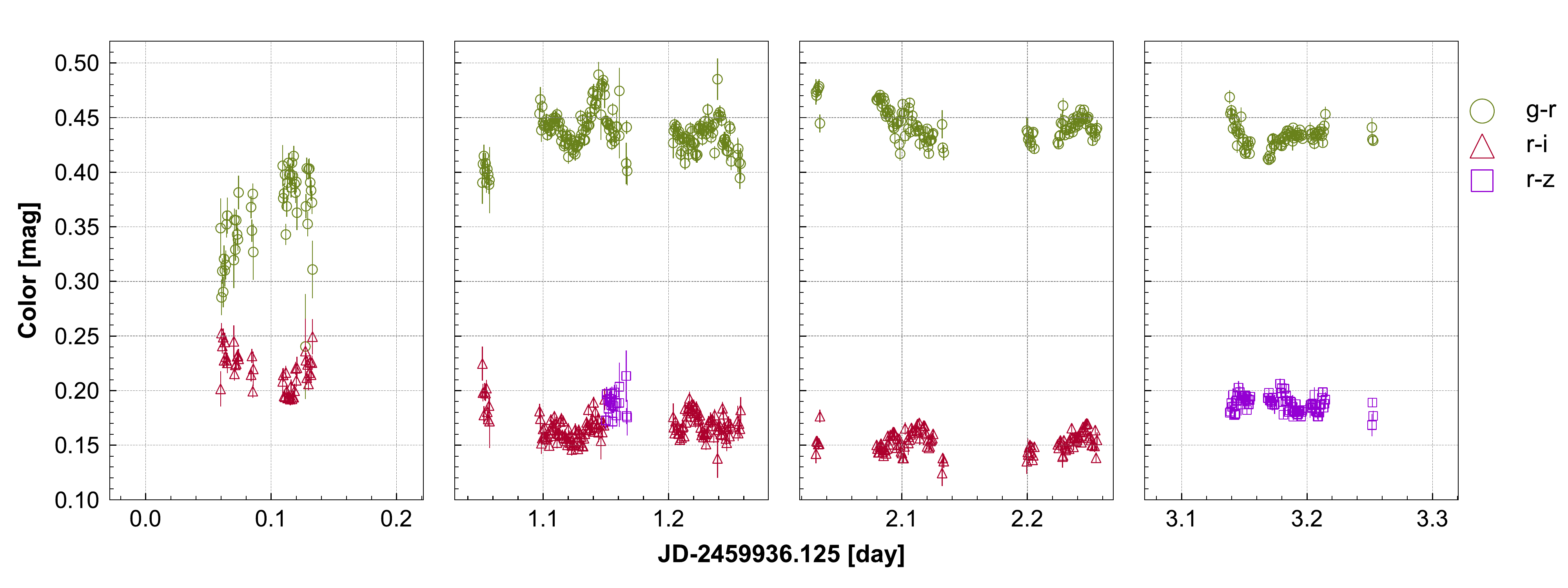}
\caption{
Time variations of colors of 2010 XC$_{15}$.
$g-r$ (circles), $r-i$ (triangles), and $r-z$ (squares) colors are presented.
Time zero is set to JD 2459936.125 (2022 December 22 15:00:00 UT).
Bars indicate the 1$\sigma$ uncertainties.
}
\label{fig:col}
\end{figure*}

We presented the colors of 2010 XC$_{15}$ in Figure \ref{fig:col}.
We estimated the systematic uncertainties of $g-r$, $r-i$, and $r-z$
colors in our observations,
$\delta_{g-r}$, $\delta_{r-i}$, and $\delta_{r-z}$, 
on the basis of the photometric measurements of reference stars \citep{Beniyama2023a}:
$\delta_{g-r}=0.03$ and $\delta_{r-i}=0.03$ in the observations with $g$, $r$, and $i$-band filters on December 23,
$\delta_{g-r}=0.02$ and $\delta_{r-z}=0.02$ in the observations with $g$, $r$, and $z$-band filters on December 23,
$\delta_{g-r}=0.02$ and $\delta_{r-i}=0.02$ in the observations with $g$, $r$, and $i$-band filters on December 24,
whereas 
$\delta_{g-r}=0.01$ and $\delta_{r-z}=0.01$ for the observations with $g$, $r$, and $z$-band filters on December 25.
The weighted average colors of 2010 XC$_{15}$, except for the first night when the condition of the sky was not ideal, 
were derived as $g-r=0.435\pm0.008$, $r-i=0.158\pm0.017$, and $r-z=0.186\pm0.009$.
The colors correspond to $V-R=0.41\pm0.02$ and $R-I=0.39\pm0.03$ in the Johnson system \citep{Tonry2012}.
We could not find notable rotational spectral variations when assuming 
the rotation periods to be 2.673~hr.

\begin{figure}[ht]
\plotone{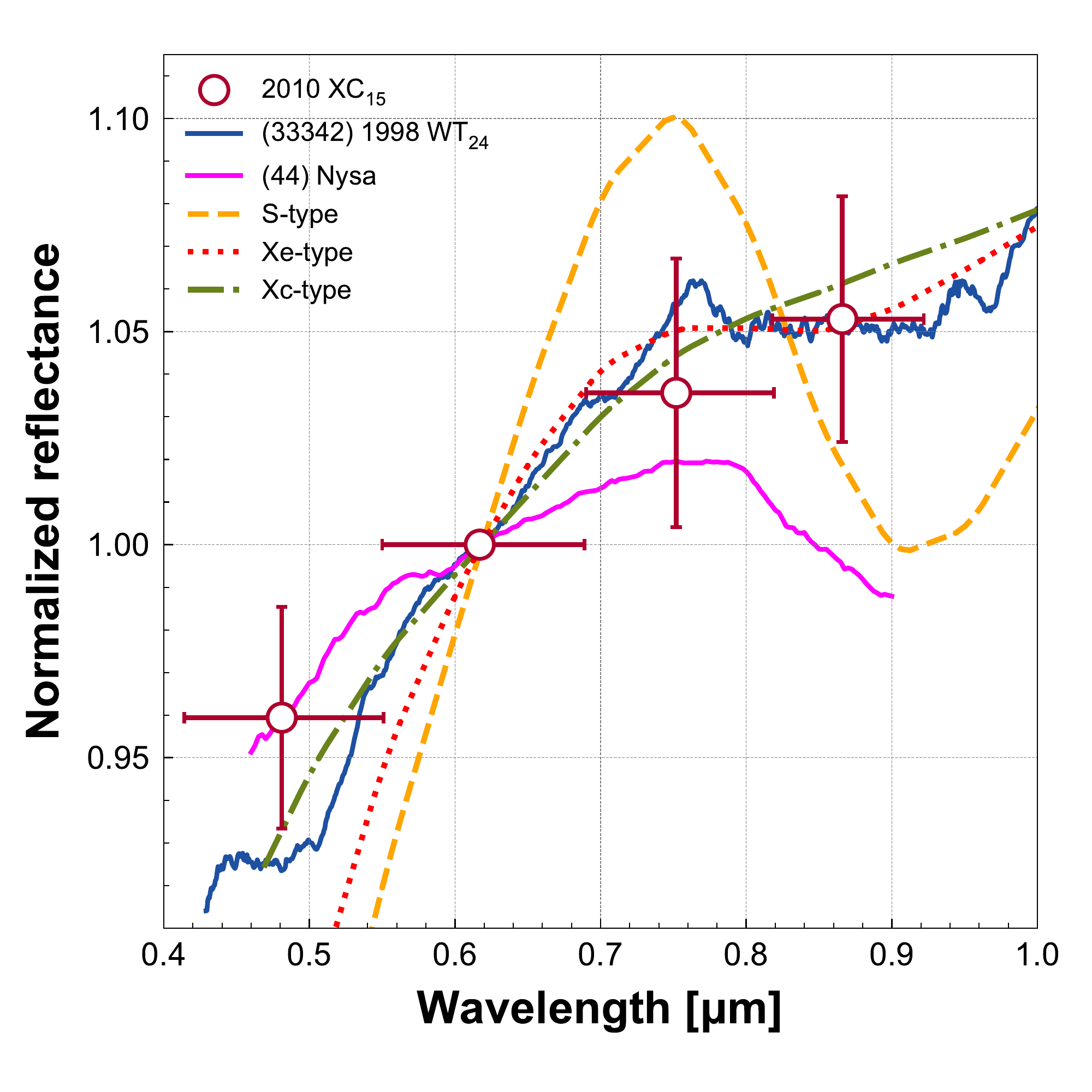}
\caption{
Reflectance spectrum of 2010 XC$_{15}$ (circles).
Vertical bars indicate the 1$\sigma$ uncertainties.
Horizontal bars indicate the filter bandwidths.
Visible spectra of other E-types are shown:
1998 WT$_{24}$ \citep[upper solid line on the right side,][]{Lazzarin2004}
and
Nysa \citep[lower solid line on the right side,][]{Bus2002b}.
Template spectra of S- (dashed line), Xe- (dotted line), and Xc-types (dot-dashed line)
are shown \citep{Bus2002b, DeMeo2013}.
The reflectance spectra 
are normalized at 0.617~$\mu$m.
}
\label{fig:ref}
\end{figure}

In Figure \ref{fig:ref}, we show the observed reflectance spectrum of 2010 XC$_{15}$. 
The reflectances at the central wavelength of the $g$, $i$, and $z$-bands, 
$R_g$, $R_i$, and $R_z$, were calculated as follows \citep[e.g.,][]{DeMeo2013}:
\begin{eqnarray}
    R_g &= 10^{-0.4[(g-r)_{\mathrm{XC_{15}}}-(g-r)_\odot]}, \\
    R_i &= 10^{-0.4[(i-r)_{\mathrm{XC_{15}}}-(i-r)_\odot]}, \\
    R_z &= 10^{-0.4[(z-r)_{\mathrm{XC_{15}}}-(z-r)_\odot]}, 
\end{eqnarray}
where 
$(g-r)_\mathrm{XC_{15}}$, $(i-r)_\mathrm{XC_{15}}$, and $(z-r)_\mathrm{XC_{15}}$ 
are colors of 2010 XC$_{15}$, 
whereas
$(g-r)_\odot$, $(i-r)_\odot$, and $(z-r)_\odot$ 
are colors of the Sun in the Pan-STARRS system.
We referred to the absolute magnitude of the Sun in the Pan-STARRS system as
$g = 5.03$, $r=4.64$, $i=4.52$, and $z=4.51$ \citep{Willmer2018}.
We set uncertainties of the magnitude of the Sun as 0.02.

The reflectance spectra in Figure \ref{fig:ref} 
    are normalized at the band center of $r$-band in the Pan-STARRS system, 0.617~$\mu$m \citep{Tonry2012}.
    Horizontal bars in 2010 XC$_{15}$'s spectrum indicate filter bandwidths \citep{Tonry2012}.
    The reflectance spectra other than 2010 XC$_{15}$ are originally normalized at 0.55~$\mu$m.
    We renormalize the spectra at 0.617~$\mu$m as follows:
    \begin{equation}
    R(\lambda)^{\prime} = \frac{R(\lambda)}{R(0.617\,\mu m)},
\end{equation}
where $R(\lambda)^{\prime}$ is a renormalized reflectance at wavelength of $\lambda$, 
$R(\lambda)$ is an original reflectance at wavelength of $\lambda$,
and $R(0.617\,\mu m)$ is an original reflectance at wavelength of 0.617~$\mu$m.
We normalized spectra after smoothing every 0.03~$\mu$m.

\subsection{Linear polarization degrees} \label{subsec:pol}
\begin{figure}[ht]
\plotone{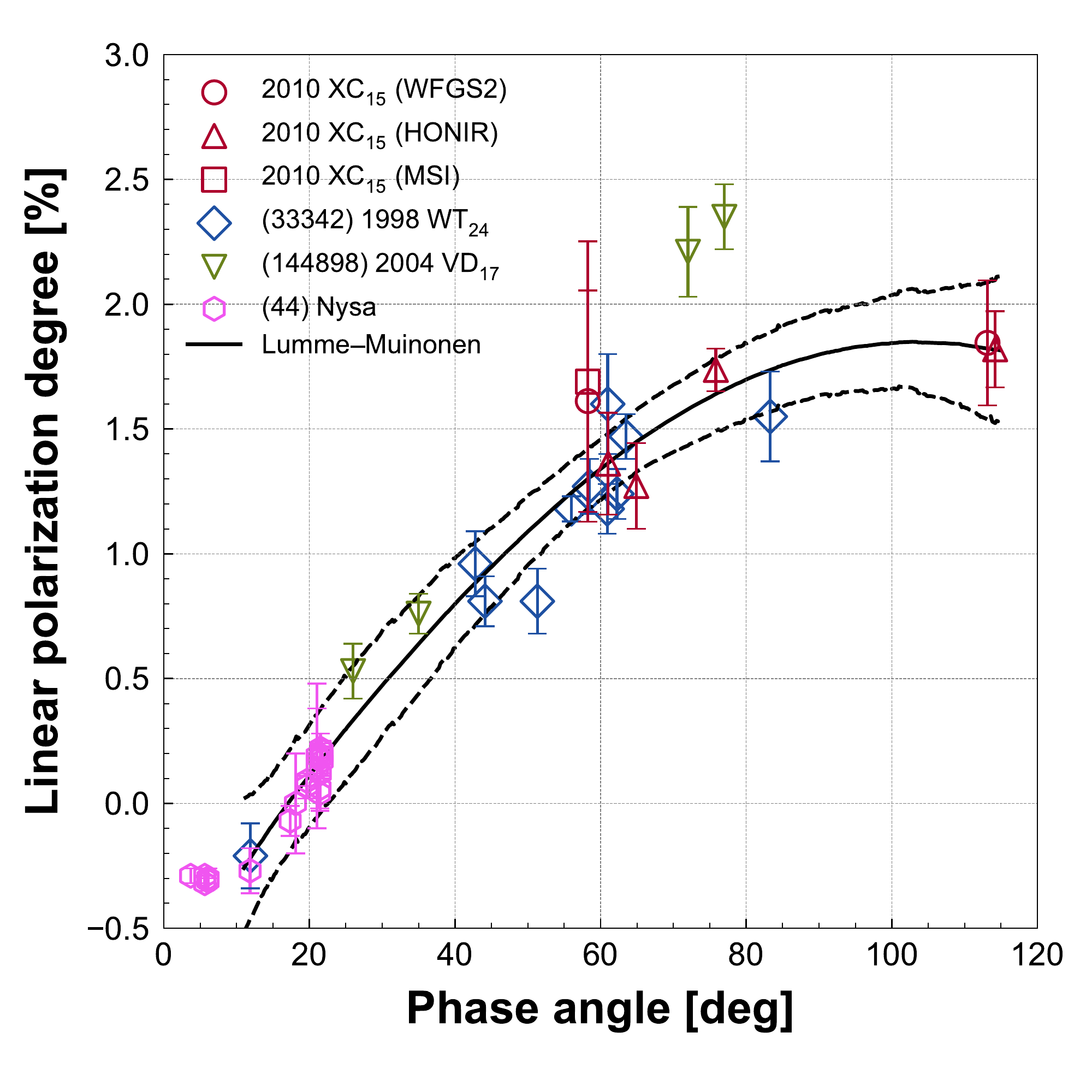}
\caption{
    Phase angle dependences of linear polarization degrees 
    of 2010 XC$_{15}$ and E-type asteroids.
    Polarization degrees of 2010 XC$_{15}$ are presented by circles (WFGS2) 
    triangles (HONIR), and a square (MSI).
    Polarization degrees of E-types are shown:
    1998 WT$_{24}$ \citep[diamonds,][]{Kiselev2002},
    2004 VD$_{17}$ \citep[inverted triangles,][]{DeLuise2007},
    and
    Nysa \citep[hexagons,][]{Zellner1976}.
    Bars indicate the 1$\sigma$ uncertainties.
    The polarization phase curve of 1998 WT$_{24}$ and 2010 XC$_{15}$
    fitted with the empirical function \citep{Lumme1993, Penttila2005} is indicated by a solid line.
    Uncertainty envelopes representing the 95 \% 
    highest density inverval (HDI) values are indicated by dashed lines.
}
\label{fig:pol}
\end{figure}

The derived linear polarization degrees and position angles of 2010 XC$_{15}$ are listed in Table \ref{tab:pol}.
Figure \ref{fig:pol} shows the observed phase angle dependence of the linear polarization degrees of 2010 XC$_{15}$.
The derived polarization degrees are a few percent at the phase angle range of 58$\degdeg$ to 114$\degdeg$.
The small linear polarization degrees 
\textbf{combined with the spectrum}
imply that 2010 XC$_{15}$ is an E-type asteroid with a high geometric albedo.
This is consistent with the $p_\mathrm{V}$ of $0.350^{+0.176}_{-0.151}$ derived from thermal observations using the IRAC on SST.

We fit the linear polarization degrees of 2010 XC$_{15}$ using an empirical model curve as follows:
\begin{eqnarray}
    P_\mathrm{r}(\alpha) = 
    b \sin^{c_1}(\alpha) \cos^{c_2}\left(\frac{\alpha}{2}\right)\sin(\alpha-\alpha_0),
\end{eqnarray}
where $b$, $c_1$, $c_2$, and $\alpha_0$ are free parameters \citep{Lumme1993, Penttila2005}.
We used the \texttt{curve\_fit} function in the \texttt{Python} package \texttt{SciPy} \citep{Virtanen2020}.
The \texttt{curve\_fit} routine determines the best fit parameters using the Levenberg–Marquardt algorithm.
Our polarimetric measurements were obtained at the phase angle range of 58$\degdeg$ to 114$\degdeg$.
The free parameters were not constrained well with only our measurements since our phase angle coverage was insufficient.
It is known that asteroids that belong to the same spectral class show a similar phase angle dependence of linear polarization 
\citep{Belskaya2017}.
As for E-types, polarimetric measurements at phase angles larger than 50$\degdeg$ have been 
reported for two NEAs:
1998 WT$_{24}$ \citep{Kiselev2002} and (144898) 2004 VD$_{17}$ \citep{DeLuise2007}.
In Figure \ref{fig:pol}, we plot the polarization degrees of these two asteroids.
These two asteroids were also not observed at small phase angles.
We plotted the polarization degrees of the prototype E-type MBA (44) Nysa \citep{Zellner1976}.
We adopted the polarization degrees determined using the $R$-band or 678 nm filter for 1998 WT24,
and those determined using the $R$-band filter for Nysa 
as our observations of 2010 XC$_{15}$ were conducted in the $R$-band.
We note that 2004 VD$_{17}$ was observed with the $V$-filter.
A good match was found between phase angle dependences of linear polarization degrees of
2010 XC$_{15}$ and 1998 WT$_{24}$, whereas the linear polarization degrees of 2004 VD$_{17}$ 
indicated a different trend.
Therefore, we combined and fit the linear polarization degrees of 2010 XC$_{15}$ and 1998 WT$_{24}$ with the empirical model curve above.
There is a good match between the model curve and linear polarization degrees of Nysa.

We generated 3000 polarization data sets by randomly resampling the measured data
assuming each polarization degree follows a normal distribution with a standard deviation of its uncertainty.
We obtained 3000 sets of fitting parameters from the generated data sets.
The maximum polarization degree of 2010 XC$_{15}$ was derived as approximately 2~\%
at $\alpha$ of approximately 100$\degdeg$.

\subsection{Dynamical evolution}\label{res:dynevo}
\begin{figure*}[ht]
\plotone{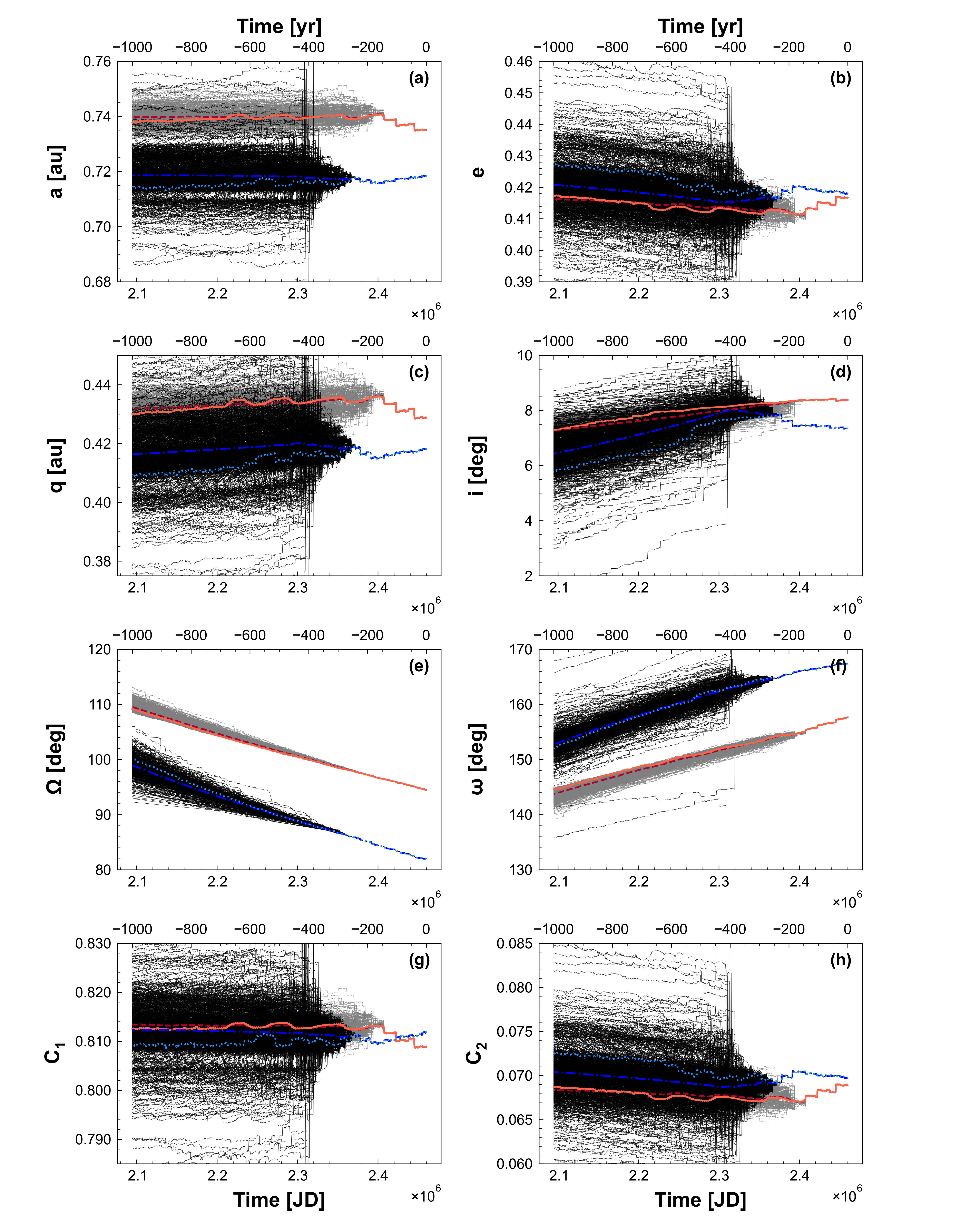}
\caption{
     Time evolution of orbital elements and integrals of 2010 XC$_{15}$ and 1998 WT$_{24}$ during 1000 years backward integration:
     (a) semimajor axis, (b) eccentricity, 
     (c) perihelion distance, (d) inclination, 
     (e) longitude of ascending node, (f) argument of perihelion,
     (g) $C_1$ integral, and (h) $C_2$ integral.
     Time evolution of nominal and averaged values for 2010 XC$_{15}$ are presented by solid and dashed lines, respectively.
     Time evolution of nominal and averaged values for 1998 WT$_{24}$ are presented by dotted and dot-dashed lines, respectively.
     Time evolution of other clones for 2010 XC$_{15}$ and 1998 WT$_{24}$ are shown by gray and black lines, respectively.
}
\label{fig:orbsim1000}
\end{figure*}

The time evolutions of the orbital elements of 2010 XC$_{15}$ and 1998 WT$_{24}$
during the last 1000 years are presented in Figure \ref{fig:orbsim1000}.
The orbital elements of 2010 XC$_{15}$ and 1998 WT$_{24}$
can be successfully traced for approximately 200 and 250 years, respectively.
The orbital elements of clones become scattered,
and chaotic behaviors can be seen.

We checked the distance between the inner planets and both 2010 XC$_{15}$ and 1998 WT$_{24}$,
outputting the coordinates and velocities of asteroids every 0.1 days.
The close approaches at a few lunar distances from the Earth-Moon system
had strong influences on the orbital evolution of both 1998 WT$_{24}$ and 2010 XC$_{15}$.
The three integrals of motion, $C_0(=1/a)$, $C_1$, and $C_2$ 
have been stable for 1000 years.
The $C_0$ of 2010 XC$_{15}$ and 1998 WT$_{24}$ 
are within the ranges of 1.350 to 1.361 and 1.391 to 1.398, respectively;
the $C_1$ of them
are within the ranges of 0.809 to 0.813 and 0.809 to 0.812, respectively;
the $C_2$ of them are within the ranges of 0.067 to 0.069 and 0.069 to 0.071, respectively.

\section{Discussion} \label{sec:disc}
\subsection{Observational confirmation of E-type NEA pair 1998 WT$_{24}$--2010 XC$_{15}$} \label{subsec:disc1}
In Figure \ref{fig:ref}, 
we show the template spectra of S-, Xe-, and Xc- types \citep[Bus-DeMeo taxonomy,][]{Bus2002b, DeMeo2013}
\footnote{\url{http://smass.mit.edu/busdemeoclass.html}},
spectrum of the E- (Tholen taxonomy) and Xc-type (Bus-DeMeo taxonomy) MBA Nysa \citep{Bus2002b},
and the E-type NEA 1998 WT$_{24}$ \citep{Lazzarin2004}.
The latter two were obtained via the M4AST online tool \citep[Modeling for Asteroids,][]{Popescu2012}.
We confirm the similarity between spectra of 2010 XC$_{15}$ and 1998 WT$_{24}$.

We check the difference of phase angles at the observations as
it is known that the slope of the visible spectrum changes depending on solar phase angles,
also known as the phase reddening effect \citep{Sanchez2012}.
Phase angles, specific dates and times of observations of 1998 WT$_{24}$ are not described in \cite{Lazzarin2004}.
We carefully combine the available information in \cite{Lazzarin2004}:
their observations were performed on 2000 October 26--28 or 2001 November 17--20
and $V$-band magnitude of 1998 WT$_{24}$ was 16.0 mag at the observations.
According to the ephemerides provided by NASA JPL HORIZONS,
the $V$-band magnitude are approximately 20 mag and 16 mag on 2000 October 26--28 and 2001 November 17--20, respectively.
The $V$-band magnitude indicates that the observations were conducted in the latter period.
The phase angles of 1998 WT$_{24}$ 
in 2001 November 17--20 were 73--77$\degdeg$,
which is not far from those of our 2010 XC$_{15}$'s observations, 58--65$\degdeg$.
Therefore, we ignore the phase reddening effect.

We derived the linear polarization degrees of 2010 XC$_{15}$ are a few percent at the phase angles across wide ranges.
The linear polarization degrees of 2010 XC$_{15}$ and 1998 WT$_{24}$
at phase angles around 60$\degdeg$ match well.
The $P_\mathrm{max}$ of 2010 XC$_{15}$ is about 2~\%.
This is almost equivalent to the $P_\mathrm{max}$ of 1998 WT$_{24}$, $1.6$--$1.8$,
derived in \cite{Kiselev2002}, 
although it must be noted that the $P_\mathrm{max}$ of 2010 XC$_{15}$ was 
derived with the polarization degrees of 1998 WT$_{24}$.
We further note that the 
$P_\mathrm{max}$ of 1998 WT$_{24}$ in \cite{Kiselev2002} 
was derived with the polarization degrees of other E-types, Nysa and (64) Angelina.

The physical properties and orbital elements of 2010 XC$_{15}$ and 1998 WT$_{24}$ are summarized in Table \ref{tab:comp}.
On the basis of the photometric and polarimetric properties described above, 
the surface properties of two E-type asteroids 2010 XC$_{15}$ and 1998 WT$_{24}$ resemble each other.
The recent spectroscopic survey of NEAs show 
E-types comprise only a few percent of the total NEA population \citep{Marsset2022a}.
Taking the similarity of not only physical properties but also dynamical integrals
and the rarity of E-types in the near-Earth region into account,
we suppose that 2010 XC$_{15}$ and 1998 WT$_{24}$ are fragments from the same parent body (i.e., asteroid pair).
They are the sixth NEA pair and the first E-type NEA pair ever confirmed.
The next close approaches of 2010 XC$_{15}$ and 1998 WT$_{24}$ 
will take place in 2027 December with $V \leq 17$ mag and in 2029 December with $V \leq 14$ mag, respectively.
Additional spectroscopic observations in wide wavelength coverage are encouraged to investigate the common origin of the two NEAs.

\begin{deluxetable*}{lrrl}
        \tablenum{3}
        \tablecaption{Comparison of physical properties and orbital elements of 2010 XC$_{15}$ and 1998 WT$_{24}$\label{tab:comp}}
        \tablewidth{0pt}
        \tablehead{
                                & 2010 XC$_{15}$ & 1998 WT$_{24}$ & References
        }
        \decimals
        \startdata
        Absolute magnitude, $H$ (mag)                        &   21.70                                   & $18.69\pm0.3$           & 1, 2  \\
        Geometric albedo, $p_\mathrm{V}$                     &   $0.350^{+0.176}_{-0.151}$                 & $0.34\pm0.20$, $0.56\pm0.2$ & 1, 2, 3\\
        Rotation period (hr)                               &   a few to several dozen          & $3.6970\pm0.0002$    & This study, 2  \\
         Volume equivalent diameter           (m)            &   $102^{+30}_{-17}$                              & $415\pm40$              & 1, 2  \\
         Maximum polarization degree, $P_\mathrm{max}$ (\%)&   $\sim2$                   & 1.6--1.8                & This study, 4  \\
         Shape                                               &   nearly spherical                               & nearly spherical               & This study, 2  \\
          Semimajor axis, $a$ (au)                            &   $  0.732375383 (0.000000015)$  & $  0.7187740919  (0.0000000027)$   & 5 \\
          Eccentricity, $e$                                   &   $  0.419862999 (0.000000040)$  & $  0.4176018439  (0.0000000098)$   & 5 \\
          Inclination, $i$ ($\degdeg$)                        &   $  8.2392588   (0.0000060)$    & $  7.3675902     (0.0000019)$      & 5 \\
          Longitude of ascending node, $\Omega$ ($\degdeg$)   &   $ 94.4069555   (0.0000035)$    & $ 81.6663922     (0.0000021)$      & 5 \\
          Argument of perihelion, $\omega$ ($\degdeg$)        &   $158.1007878   (0.0000069)$    & $167.5262827     (0.0000028)$      & 5 \\
          Mean anormaly, $M$ ($\degdeg$)                      &   $323.063632    (0.000012)$     & $136.978426      (0.000022)$       & 5 \\
  \enddata
            \tablecomments{
            (1) NEO Legacy
            (2) \cite{Busch2008}
            (3) Best-fit values from thermal-infrared observations in \cite{Harris2007}
            (4) \cite{Kiselev2002}
            (5)
            The orbital elements of 2010 XC$_{15}$ are
            referred to epoch Julian Day 2460000.5 (2023 February 25.0)
            TDB (Barycentric Dynamical Time, J2000.0 ecliptic and equinox).
            It is based on 279 observations with a data-arc span of 4406 days
            (solution date, 2023 February 14 15:10:21). 
            The orbital elements of 1998 WT$_{24}$ are
            referred to epoch Julian Day 2460000.5 (2023 February 25.0) TDB.
            It is based on 1842 observations with a data-arc span of 8489 days
            (solution date, 2023 March 1 06:14:53). 
            Information above together with orbital elements are extracted from NASA JPL \textbf{SBDB}.
            Values in the parentheses are 1$\sigma$ uncertainties of orbital elements.
            }
            \end{deluxetable*}

\subsection{Dynamical history and origin of E-type NEA pair}\label{subsec:disc2}
\begin{figure}[ht]
\plotone{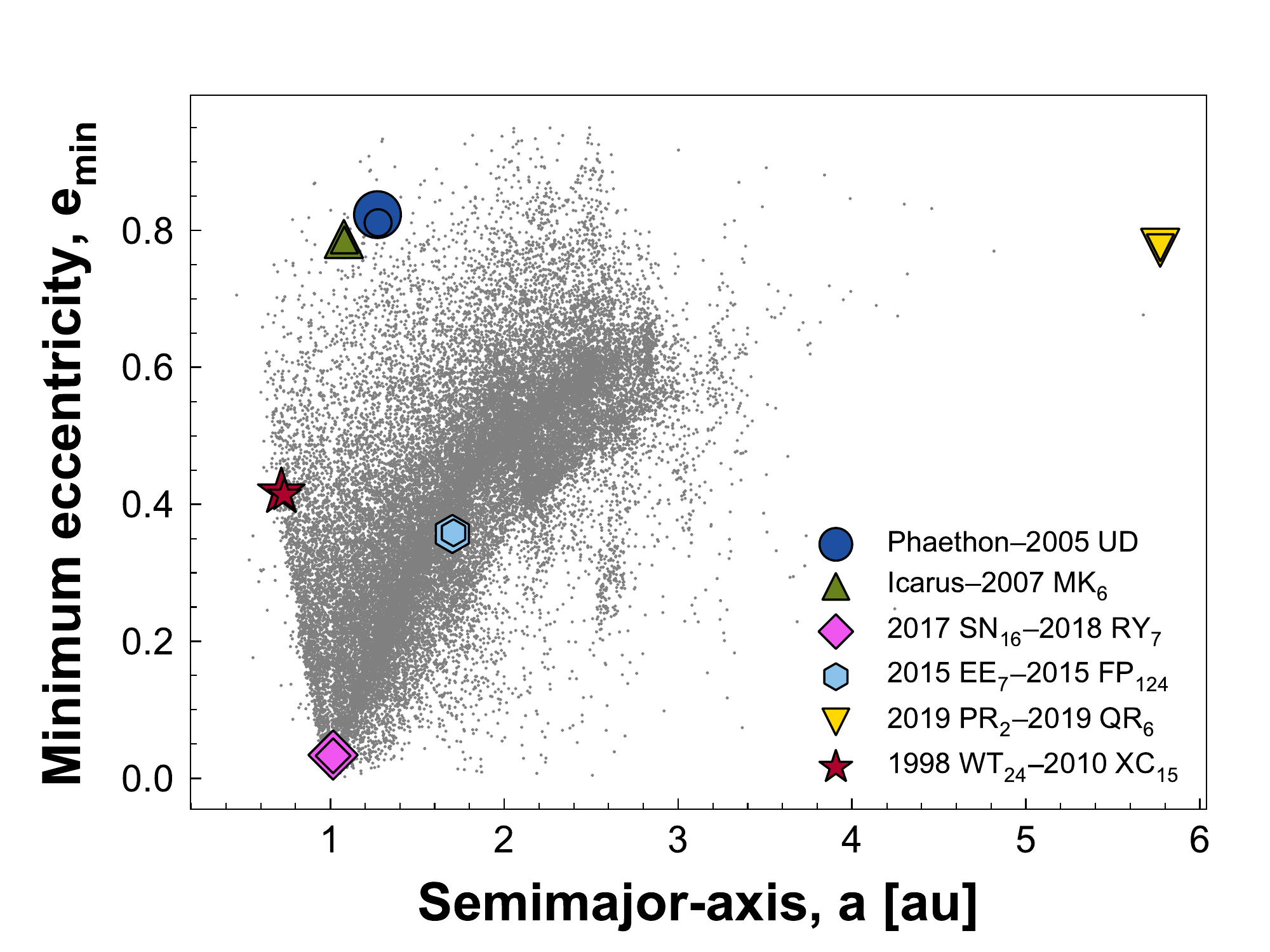}
\caption{
    Semimajor-axis ($a$) vs. minimum eccentricity ($e_\mathrm{min}$) of NEAs.
    Larger (primary) asteroids are plotted with larger markers than
    smaller (secondary) asteroids for the all pairs.
    Orbital elements are extracted from the NEODyS-2 as of 2023 May 1.
}
\label{fig:ae}
\end{figure}

\begin{figure}[ht]
\plotone{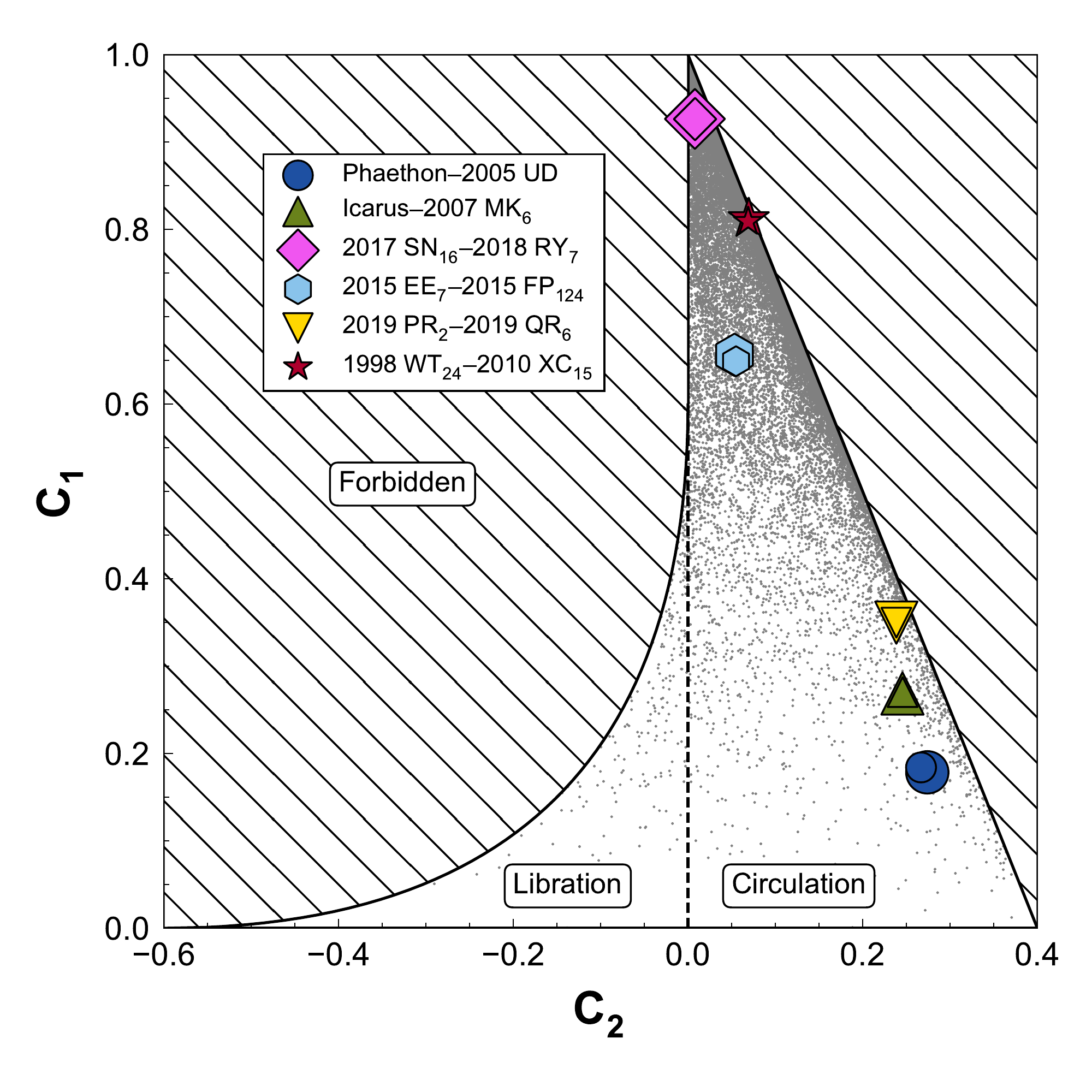}
\caption{
    Lidov diagram with $C_1$ and $C_2$ of NEAs.
    Larger asteroids (primary) are plotted with larger markers than
    smaller asteroids (secondary) for the all pairs.
    The area to the left of the separatrix (dashed line) is libration mode
    (argument of perihelion librates),
    and the area to the right is circulation mode (argument of perihelion circulates).
    The forbidden region is hatched.
    Orbital elements used to calculate $C_1$ and $C_2$ are
    extracted from the Minor Planet Center Orbit Database file
    as of 2023 May 1.
}
\label{fig:Lidov}
\end{figure}

The six confirmed NEA pairs are presented on an 
$a$--$e_\mathrm{min}$ or $1/C_0$--$e_\mathrm{min}$ plane in Figure \ref{fig:ae}:
Phaethon--2005 UD \citep{Ohtsuka2006},
Icarus--2007 MK$_6$ \citep{Ohtsuka2007},
2017 SN$_{16}$--2018 RY$_7$ \citep{Marcos2019c, Moskovitz2019},
2015 EE$_{7}$--2015 FP$_{124}$ \citep{Moskovitz2019},
2019 PR$_2$--2019 QR$_6$ \citep{Fatka2022},
and 1998 WT$_{24}$--2010 XC$_{15}$.
The $e_\mathrm{min}$ is a minimum value of the orbital eccentricity over one period of $\omega$ \citep{Gronchi2001}.
The orbital elements were extracted from the NEA element catalogs of NEODyS-2\footnote{\url{https://newton.spacedys.com/neodys/}}
as of 2023 May 1.
The $C_1$ and $C_2$ of 2010 XC$_{15}$, 1998 WT$_{24}$,
and other NEA pairs are plotted on the Lidov diagram in Figure \ref{fig:Lidov} \citep{Lidov1962, Ito2019}.
The Lidov diagram helps us understand the dynamical characteristics of the system.
The orbital elements used to calculate $C_1$ and $C_2$ were
extracted from the Minor Planet Center Orbit Database file\footnote{\url{https://minorplanetcenter.net/iau/MPCORB/NEA.txt}}
as of 2023 May 1.
The separations of 2010 XC$_{15}$ and 1998 WT$_{24}$ on Figures \ref{fig:ae} and \ref{fig:Lidov} 
are as small as the other NEA pairs as summarized in Table \ref{tab:pairs}.
This supports the idea that the two asteroids are of common origin 
as these integrals are useful indicators to confirm NEA pairs
\citep{Ohtsuka2006, Ohtsuka2007}.
As shown in Figure \ref{fig:orbsim1000},
the orbital elements of 2010 XC$_{15}$ and 1998 WT$_{24}$ become scattered
after backward integrations of 200 and 250 years, respectively.
Thus, we could not determine the exact time of the breaking up event.

It is worth mentioning that the mass ratio of 1998 WT$_{24}$--2010 XC$_{15}$
as well as Phaethon--2005 UD and Icarus--2007 MK6
is close to 0.25.
The rotation period of the primary 1998 WT$_{24}$ is 3.6970 hr \citep{Busch2008},
which is also close to that of another primary Phaethon.
\cite{Busch2008} revealed that the shape of 1998 WT$_{24}$ looks like a spherical body 
with three basins.
They interpreted that the basins may be impact craters or a relic of past dynamical disruption.
The overall shape of 1998 WT$_{24}$ resembles 
top-shaped asteroids such as 2008 EV$_5$ and 2000 DP$_{107}$ on which 
rotational fissions may have occurred \citep{Tardivel2018}.
The mass ratio and rotation period are consistent with the theory of 
the rotation fission \citep{Scheeres2007b, Pravec2010}.
This is also the case for the Phaethon--2005 UD pair \citep{Hanus2016}.
Therefore, the rotational fission is favored as the formation mechanism of the 1998 WT$_{24}$--2010 XC$_{15}$ pair.
Considering the diameter of the parent body of 1998 WT$_{24}$--2010 XC$_{15}$ is almost
equivalent to that of 1998 WT$_{24}$ at approximately 400~m,
the YORP spin-up might play an important role in rotational fission
since it strongly changes the rotation state of such small bodies \citep{Rubincam2000}.
In terms of the spherical shape of the primary 1998 WT$_{24}$,
rotational fission is preferred instead of tidal disruption,
which typically expects elongated primaries \citep{Walsh2006}.

There are some escape regions from MBAs to NEAs \citep{Granvik2018}.
One is the $\nu_6$ resonance at the inner edge of main-belt
with a semimajor axis of approximately 2.1~au.
Nysa is located at the inner main-belt with a semimajor axis of approximately 2.4~au.
\cite{Reddy2016} characterized a tiny ($D\sim2$~m) E-type NEA 2015 TC$_{25}$
in radar, optical lightcurves, and near-infrared spectroscopic observations.
They combined the spectral and dynamical properties of 2015 TC$_{25}$
and concluded that it is a fragment possibly ejected from Nysa.
Another representative escape region for E-types is the Hungaria region with a semimajor axis of approximately 1.9~au and an inclination of about 20$\degdeg$.
It is known that the fraction of NEAs from the Hungaria region is 
smaller than those from the inner main-belt through the $\nu_6$ resonance \citep{Granvik2018}. 
On the other hand, 
considering the relative fractions of E-types in different regions of the main belt, 
the probability of them originating from the Hungaria region is higher than 
from the $\nu_6$ resonance or comparable within uncertainties \citep{Binzel2019}.
Thus, the source region of 2010 XC$_{15}$, 1998 WT$_{24}$, and their parent body cannot be clearly determined.

\begin{deluxetable*}{lccccc}
        \tablenum{4}
        \tablecaption{Proper orbital elements and integrals of NEA pairs\label{tab:pairs}}
        \tablewidth{0pt}
        \tablehead{
           & $a$ & $e_\mathrm{min}$ & $C_0$ & $C_1$ & $C_2$ 
        }
        \decimals
        \startdata
        Phaethon            & 1.2713  & 0.8229   & 0.78661  & 0.17842 & 0.27397 \\
        2005 UD             & 1.2748  & 0.8111   & 0.78444  & 0.18403 & 0.26681 \\
        \it{difference}     & 0.0035  & 0.0118   & 0.00218  & 0.00561 & 0.00716 \\
        Icarus              & 1.0779  & 0.7875   & 0.92770  & 0.26871 & 0.24558 \\
        2007 MK$_{6}$       & 1.0808  & 0.7862   & 0.92521  & 0.27029 & 0.24572 \\
        \it{difference}     & 0.0029  & 0.0012   & 0.00248  & 0.00158 & 0.00015 \\
        2017 SN$_{16}$      & 1.0161  & 0.0339   & 0.98417  & 0.92635 & 0.00797 \\
        2018 RY$_{7}$       & 1.0162  & 0.0328   & 0.98403  & 0.92618 & 0.00813 \\
        \it{difference}     & 0.0001  & 0.0012   & 0.00015  & 0.00017 & 0.00016  \\
        2015 EE$_{7}$       & 1.7017  & 0.3567   & 0.58765  & 0.65636 & 0.05328  \\
        2015 FP$_{124}$     & 1.7084  & 0.3584   & 0.58533  & 0.64897 & 0.05501 \\
        \it{difference}     & 0.0067  & 0.0017   & 0.00232  & 0.00738 & 0.00172  \\
        2019 PR$_{2}$       & 5.7721  & 0.7744   & 0.17325  & 0.34974 & 0.23851\\
        2019 QR$_{6}$       & 5.7727  & 0.7745   & 0.17323  & 0.34972 & 0.23857\\
        \it{difference}     & 0.0006  & 0.0001   & 0.00002  & 0.00001 & 0.00006   \\
        1998 WT$_{24}$      & 0.7185  & 0.4179   & 1.39183  & 0.81205 & 0.06962 \\
        2010 XC$_{15}$      & 0.7349  & 0.4135   & 1.36073  & 0.80887 & 0.06889 \\
        \it{difference}     & 0.0164  & 0.0043   & 0.03110  & 0.00318 & 0.00073  
  \enddata
        \tablecomments{
            Seminajor axis, $a$, minimum eccentricity, $e_\mathrm{min}$, $C_0$,
            $C_1$, and $C_2$ of NEA pairs.
            The absolute differences of five parameters for each pair are listed for each pair.
            The $a$($\equiv1/C_0$) and $e_\mathrm{min}$ are extracted from the NEA element catalogs of NEODyS-2 as of 2023 May 1.
            Orbital elements used to calculate $C_1$ and $C_2$ are extracted from 
            the Minor Planet Center Orbit Database file as of 2023 May 1.
            }
            \end{deluxetable*}

\section{Conclusion} \label{sec:conc}
We performed optical photometry and polarimetry of a small NEA 2010 XC$_{15}$ in 2022 December.
We found that the rotation period of 2010 XC$_{15}$ is possibly a few to several dozen hours
and 
color indices of 2010 XC$_{15}$ are $g-r=0.435\pm0.008$, $r-i=0.158\pm0.017$, and $r-z=0.186\pm0.009$ in the Pan-STARRS system.
Additionally, we derived the linear polarization degrees of 2010 XC$_{15}$ is a few percent 
at the phase angle range of 58$\degdeg$ to 114$\degdeg$.
We found that 2010 XC$_{15}$ is a rare E-type NEA on the basis of its photometric and polarimetric properties. 
Taking the similarity of not only physical properties but also dynamical integrals and the rarity of E-type NEAs into account,
we suppose that 2010 XC$_{15}$ and 1998 WT$_{24}$ are an E-type NEA pair.
These two NEAs are the sixth NEA pair and first E-type NEA pair ever confirmed and were possibly formed by rotational fission.
We conjecture that the parent body of 2010 XC$_{15}$ and 1998 WT$_{24}$ 
is from the main-belt througn the $\nu_6$ resonance or the Hungaria region.
The next observing windows of 2010 XC$_{15}$ and 1998 WT$_{24}$ 
will arrive in December 2027 with $V \leq 17$ mag and December 2029 with $V \leq 14$ mag, respectively.
Additional spectroscopic observations in wide wavelength coverage are encouraged to further investigate the common origin of these two NEAs.

\begin{acknowledgments}
We would like to thank to Jooyeon Geem and Ryota Fukai 
for their help and discussions regarding polarimetry and E-type asteroids, respectively.
We are grateful for the insights provided by Patrick Michel.
The authors are grateful to the anonymous referee 
for a very constructive review of this manuscript.
J. B. would like to express gratitude to the Public Trust Iwai Hisao Memorial Tokyo Scholarship Fund for the grants.
This research is partially supported by the Optical and 
Infrared Synergetic Telescopes for Education and Research (OISTER) program funded by the MEXT of Japan.
This work is supported in part by the JST SPRING, Grant Number JPMJSP2108, and the UTEC UTokyo Scholarship
as well as a grant from the Hayakawa Satio Fund awarded by the Astronomical Society of Japan.
This work has been supported by the Japan Society for the Promotion of Science (JSPS) 
Grants-in-Aid for Scientific Research (KAKENHI) Grant Number 21H04491 and 23KJ0640.
The authors thank the TriCCS developer team (which has been supported by the JSPS KAKENHI grant Nos. JP18H05223,
JP20H00174, and JP20H04736, and by NAOJ Joint Development Research).
The Pan-STARRS1 Surveys (PS1) and the PS1 public science archive have been
made possible through contributions by the Institute for Astronomy,
the University of Hawaii, the Pan-STARRS Project Office,
the Max-Planck Society and its participating institutes,
the Max Planck Institute for Astronomy, Heidelberg and
the Max Planck Institute for Extraterrestrial Physics, Garching,
The Johns Hopkins University, Durham University, the University of Edinburgh,
the Queen's University Belfast, the Harvard-Smithsonian Center for Astrophysics,
the Las Cumbres Observatory Global Telescope Network Incorporated,
the National Central University of Taiwan, the Space Telescope Science Institute,
the National Aeronautics and Space Administration under Grant No. NNX08AR22G
issued through the Planetary Science Division of the NASA Science Mission Directorate,
the National Science Foundation Grant No. AST-1238877, the University of Maryland,
Eotvos Lorand University (ELTE), the Los Alamos National Laboratory,
and the Gordon and Betty Moore Foundation.
We wish to thank Department of Mathematics, University of Pisa,
IASF-INAF, SpaceDyS srl, JPL, Department of Applied Mathematics, University of Valladolid,
Hyperborea srl, and OrbFit Consortium
for operating and contributing a convenient web-based interface NEODyS.
\end{acknowledgments}

\vspace{5mm}
\facilities{Seimei (TriCCS), Nayuta (WFGS2), Kanata (HONIR), Pirka (MSI)}

\software{
NumPy \citep{Oliphant2015, Harris2020},
pandas \citep{McKinney2010},
SciPy \citep{Virtanen2020},
AstroPy \citep{Astropy2013, Astropy2018},
Astro-SCRAPPY \citep{McCully2018},
astroquery \citep{Ginsburg2019}, 
Matplotlib \citep{Hunter2007},
Source Extractor \citep{Bertin1996},
SEP \citep{Barbary2015},
astrometry.net \citep{Lang2010}
}

\appendix
\section{Validation of polarimetric measurements} \label{app:polstan}
\begin{figure}\epsscale{0.8}
    \plotone{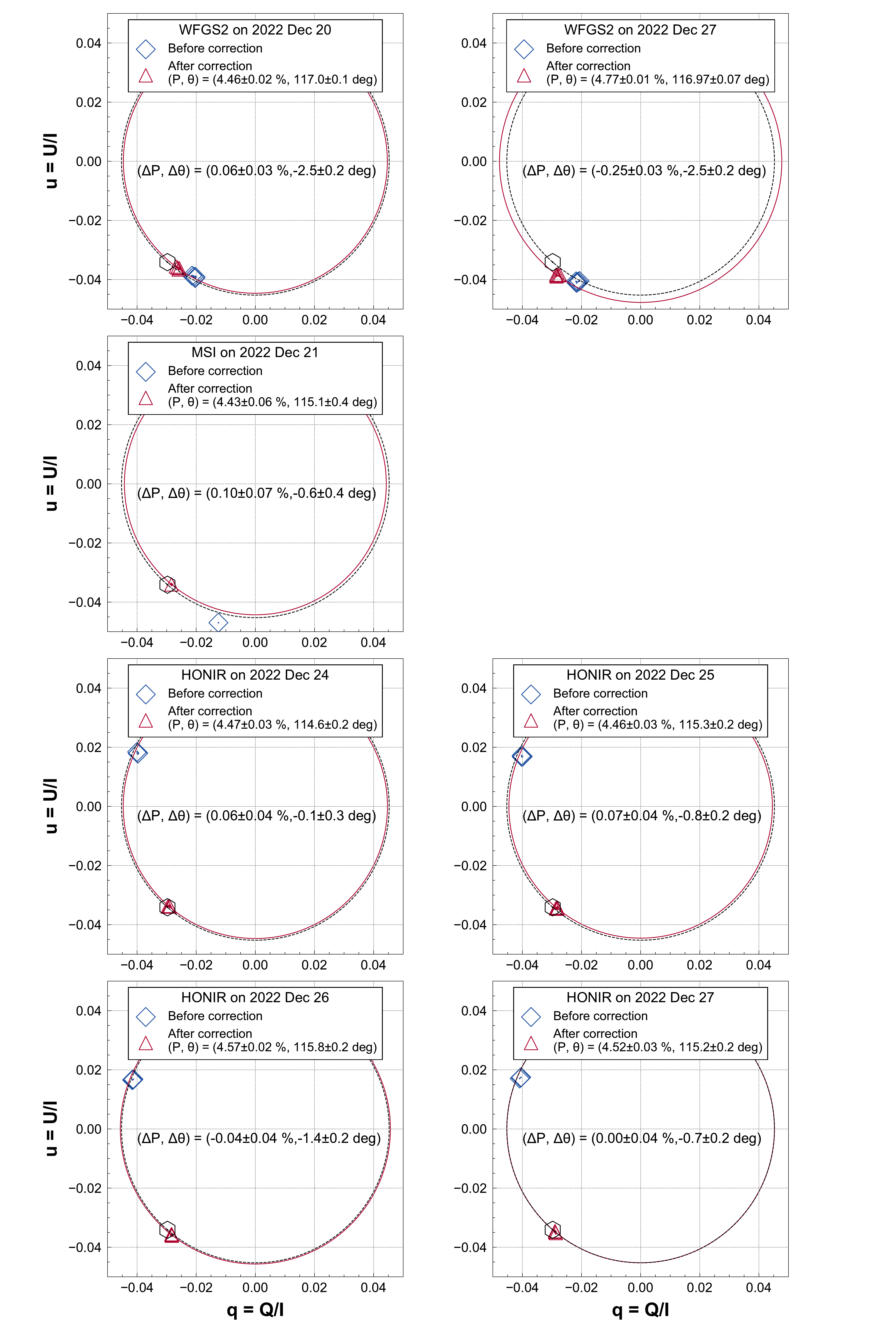}
\caption{
Results of polarimetry of polarimetric standard star HD 19820 at each site on each day.
The Stokes parameters $Q$ and $U$ normalized by the intensity $I$,
$q\equiv Q/I$ and $u\equiv U/I$, before and after series of corrections 
are presented as diamonds and triangles, respectively.
The $q$ and $u$ of HD 19920 calculated from $P$ and $\theta$ in \cite{Schmidt1992} 
are shown by hexagons.
Bars indicate the 1$\sigma$ uncertainties.
Polarization degrees of HD 19820 after series of corrections and 
those in \cite{Schmidt1992} are presented by large solid circles and dashed circles, respectively.
The derivations $\Delta P$, 
polarization degrees in \cite{Schmidt1992} - those derived here,
and $\Delta \theta$,
position angles of polarization in \cite{Schmidt1992} - those derived here,
are shown.
}
\label{fig:HD19820}
\end{figure}

We present the polarimetric results of a polarimetric standard star HD 19820 
\citep{Schmidt1992} in Figure \ref{fig:HD19820} for validation purposes.
We derived the linear polarization degrees, $P$, 
and position angles of polarization, $\theta$, of HD 19820 with high accuracy.
We considered the derivations between the photometric parameters 
in \cite{Schmidt1992} and those derived here 
as systematic uncertainties in the measurements of the polarimetric parameters of 2010 XC$_{15}$.

\newcommand{\noopsort}[1]{} \newcommand{\singleletter}[1]{#1}

\bibliographystyle{aasjournal}

\end{document}